\documentclass[journal]{IEEEtran}

\ifCLASSINFOpdf
\else
   \usepackage[dvips]{graphicx}
\fi
\usepackage{url}
\usepackage{float}
\usepackage{svg}
\usepackage[utf8]{inputenc}
\usepackage{pdfpages}
\usepackage{graphicx}
\usepackage{wrapfig}
\usepackage{multicol}
\usepackage{svg}
\usepackage{MnSymbol}
\usepackage{hyperref}
\usepackage{amsfonts} 
\usepackage{booktabs}
\usepackage{tikz}
\usepackage{pgfplots}
\pgfplotsset{width=5.5cm,compat=1.12}
\usepgfplotslibrary{fillbetween}
\usepackage[skip=4pt plus1pt, indent=8pt]{parskip}
\newcommand\linesubsec[1]{\vspace{0.8mm}\noindent\textbf{#1 --- }}
\usepackage{hyperref}

\begin{document}

\title{ATGNN: Audio Tagging Graph Neural Network}

\author{Shubhr Singh, Christian J. Steinmetz, Emmanouil Benetos, \IEEEmembership{Sr. Member, IEEE}, Huy Phan and Dan Stowell
%\thanks{This paragraph of the first footnote will contain the date on which you submitted your paper for review. It will also contain support information, including sponsor and financial support acknowledgment.}
\thanks{SS and CS are supported jointly by UK Research and Innovation  and Queen Mary University of London under grant EP/S022694/1. This work was supported by the Engineering and Physical Sciences Research Council [grant number EP/V062107/1].}
\thanks{SS, CJ and EB  are with the School of Electronic Engineering and Computer Science, Queen Mary University of London, UK (e-mail:
\{\texttt{s.shubhr, c.j.steinmetz, emmanouil.benetos\}@qmul.ac.uk}). HP is with Amazon Alexa, Cambridge, MA 02142, USA (e-mail: \texttt{huypq@amazon.co.uk}). DS is with Tilburg University, Bijsterveldenlaan, 5037 AB Tilburg, Netherlands (e-mail: \texttt{d.stowell@tilburguniversity.edu}).}
\thanks{The work was done when HP was at the School of Electronic Engineering and Computer Science, Queen Mary University of London, UK and prior to joining Amazon.}
}

\markboth{IEEE SIGNAL PROCESSING LETTERS, VOL. XX, 2023}
{Shell \MakeLowercase{\textit{et al.}}: Bare Demo of IEEEtran.cls for IEEE Journals}
\maketitle

\begin{abstract}
Deep learning models such as CNNs and Transformers have achieved impressive performance for end-to-end audio tagging. Recent works  have shown that despite stacking multiple layers, the receptive field of CNNs remains severely limited. Transformers  on the other hand are able to map global context through self-attention, but treat the spectrogram as a sequence of patches which is not flexible enough to capture irregular audio objects. In this work, we treat the spectrogram in a more flexible way by considering it as graph structure and process it with a novel  graph neural architecture called ATGNN. ATGNN not only combines the capability of CNNs with the global information sharing ability of Graph Neural Networks, but also maps semantic relationships between learnable class embeddings and corresponding spectrogram regions. We evaluate ATGNN on two audio tagging tasks, where it achieves 0.585 mAP on the FSD50K dataset and 0.335 mAP on the AudioSet-balanced dataset, achieving comparable results to Transformer based models with significantly lower number of learnable parameters.
\end{abstract}


\IEEEpeerreviewmaketitle

\section{Introduction}

Environmental sounds carry a rich and complex mixture of
information, which is organised and categorised by the human auditory system into distinct concepts known as \textit{sound events } (e.g. dog bark, door slam, car passing by). The advent of deep learning models such as convolutional neural networks (CNNs) revolutionised the research field of image classification and  achieved state of the art on numerous audio classification \& tagging benchmarks as well~\cite{hershey2017cnn}.
%However, unlike objects within images, sound events often exhibit non-locality within the spectrogram representation. This is a result of the reality that sound events are often composed of harmonic components that resonate at non-local frequencies known as \textit{signal harmonics}.   

%Therefore, in order to perform well in tasks like audio tagging where multiple sound events are present, models must be capable of detecting long range as well as short range dependencies across both the temporal and frequency axes. However, even when stacking multiple convolution layers, the effective receptive field of these models remains limited, essentially impeding their ability to map global context~\cite{luo2016understanding}.

In recent times, Transformer-based models~\cite{ast,koutini2021efficient} have demonstrated superior performance over CNNs in audio tagging and classification tasks. This can be attributed to their unique capability to capture global context by employing self-attention mechanisms, treating spectrograms as sequences of patches. Despite the appealing advantages of Transformers, their usage comes with a trade-off as they tend to incur substantial computational costs during both training and inference stages~\cite{liu2023efficientvit}.

%Recently, Transformer based  models~\cite{ast,koutini2021efficient}  were able to outperform CNNs in audio tagging \& classification tasks due to their ability to map global context using self-attention mechanism by treating spectrogram  as a sequence of  patches. While Transformers are appealing, they often incur large computational costs during training and inference.  

Sound events can be conceptualized as  compositions of multiple regions in the spectrogram that are interlinked to form a coherent representation of the event. This underlying structure naturally lends itself to a graph-based representation, where each region in the spectrogram becomes a node, and the relationships between these regions are captured through edges. Graphs provide a flexible structure as compared to grid or sequences of patches and also open the potential for harnessing label correlation information, enhancing the overall predictive capabilities. Label co-occurrence graphs (LG)~\cite{chen2019multi} represent the relationships and co-occurrence patterns among different labels in a dataset. LG has been used to improve the performance of models in multi-label image recognition~\cite{chen2019multi} as well as in audio tagging~\cite{shrivastava2020mt}. Keeping the benefits of graph structure in mind and given the recent success of graph based models in image classification~\cite{VIG-GNN,yaoml}, we propose the \textit{Audio Tagging graph neural network} (ATGNN), an end-to-end graph convolution network for audio tagging applications. 

In our approach, we begin by extracting features from an input spectrogram using a backbone CNN. Each element of the resulting feature map is then considered as an individual node. To enable effective communication between distant patches in the original spectrogram, we dynamically construct edges between nodes based on their similarity in the feature space, facilitating  the exchange of information across different regions of the spectrogram, allowing our model to capture and leverage complex dependencies. 

In addition to employing graph processing in the feature space, our approach incorporates learnable label embeddings, serving a dual purpose in capturing two distinct relationships. Firstly, these embeddings facilitate the modelling of semantic relationships across class labels, enabling the model to comprehend the underlying correlations between different sound event categories. Secondly, the label embeddings establish cross-domain relationships between spectrogram regions and label embeddings, establishing connections between the labels and regions of interest in the spectrogram, enhancing the overall discriminative capability of our model.
%The graph is processed the node paireach node(label)is representedbywordembeddingsofalabel has been shown to improve the performance of models in audio tagging of using label correlation information\textbf{}
%While pure attention based models are capable of outperforming CNNs in audio classification tasks, they often require more compute during training and inference. 

%One appealing idea is to combine the ability of CNNs to capture local context with models that can map global context such as graphs, which have been used in the image domain for context modelling and reasoning with local image regions as nodes and similarity between them as edges~\cite{beyond1,li2018beyond}.
%Given the recent success of graph based models in image classification~\cite{VIG-GNN,yaoml} and shortcomings of standard CNNs in time-frequency domain, we introduce \textit{Spectrogram GNN} (SpecGNN), an end to end graph convolution model, which learns a graph representation  on the  2D feature maps of a backbone CNN.  

%Nodes within the graph are local patches of the input spectrogram represented by the values of the feature map. Edges are constructed dynamically based on similarity of the nodes in the feature space, which enables communication between distance patches in the original spectrogram. The model explicitly learns spatial relationship between local spectrogram regions along with semantic relationship between class labels and cross-relationship  between spectrogram regions and class labels in an end-end manner. 

There has been a steady increase in the adoption of GNNs for audio tasks such as speech emotion recognition (SER)~\cite{liu2021time}, speaker diarisation~\cite{kwon2022multi,wang2020speaker}, audio tagging~\cite{sun2020ontology, shrivastava2020mt}. 

Our research is different from prior works as it leverages spectrogram graph structures to effectively utilize inter-region relationships, and applies an end-to-end approach to concurrently learn label correlations and label-spectrogram interactions.

%Our research distinguishes itself from prior works in two significant aspects. Firstly, we treat the spectrogram as a graph structure, allowing us to leverage the inherent relationships between spectrogram regions effectively. Secondly, our model adopts an end-to-end approach to simultaneously learn label correlations and label-spectrogram relationships. 
%In the case of environmental sound classification, graph neural networks (GNN) have been primarily employed for modelling label correlation and fused with a CNN based model in order to inject domain knowledge about the label structure into the model 
%Our work differs from previous graph based approaches as it neither uses label embeddings extracted from a language model, nor does it use a pre-defined graph structure. Both the entities are learnt in an end-to-end fashion. The closest work to ours is VIG-GNN~\cite{VIG-GNN} and ML-VIG~\cite{yaoml}, both of which are an end-end graph models for the image domain and serve as an inspiration for our work. 

The main contribution of this paper is therefore summarised as follows:
(i) We model the spectrogram as a graph structure and propose an end-to-end graph based model for audio tagging. (ii) To the best of our knowledge, this is the first attempt to simultaneously model correlations between labels and spectrograms, as well as between labels themselves, in an end-to-end fashion without relying on previously established correlations and demonstrates the effectiveness of the methodology on two widely used audio tagging datasets.

\begin{figure*}[t]
  \centering
  %\includesvg[scale =0.1]{Interspeech_5.svg}
  \includegraphics[width = 0.8\textwidth]{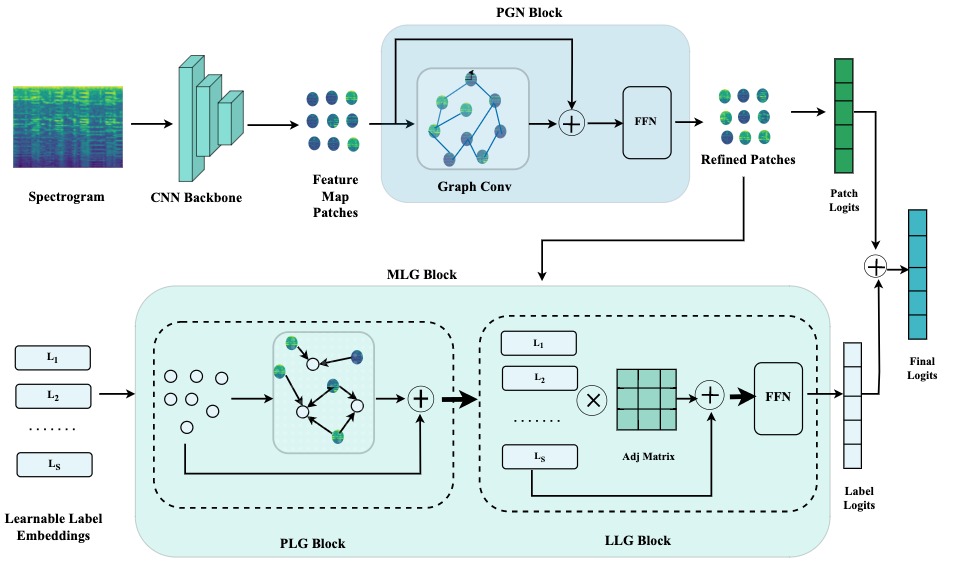}
  \caption{\textbf{ATGNN}: The input spectrogram is passed through a CNN backbone and  a k-nearest neighbour graph is constructed with the the feature map pixel as nodes. The patch nodes are updated with graph convolution across $M$ PGN  blocks and then fed into an MLG block where cross-correlation between learnable label embeddings and patch nodes is learnt. Each label node connects to its nearest patch nodes and is updated using graph convolution. The LLG block maps label-label correlation using a learned adjacency matrix.  A stack of PGN and MLG blocks is used to refine the patch and label embeddings and is combined for final prediction.}
\end{figure*}

%\section{Related Work}
%Although graph based models are primarily employed for modelling non Euclidean data, there has been a steady increase in adoption of either graph structured data for speech domain such as speech emotion recognition (SER)~\cite{liu2021time,lian2020conversational} and speaker diarisation~\cite{kwon2022multi,wang2020speaker}. In the case of environmental sound classification, graph neural networks (GNN) have been primarily employed for  modelling label correlation and fused with a CNN based model in order to inject domain knowledge about the label structure into the model~\cite{sun2020ontology, shrivastava2020mt}. Our work is significantly different from previous graph based approaches as it neither uses label embeddings extracted from a language model, nor does it use a pre-defined graph structure. Both the entities are learnt in an end-to-end fashion. The closest work to ours is VIG-GNN~\cite{VIG-GNN} and ML-VIG~\cite{yaoml}, both of which are an end-end graph models for the image domain and serve as an inspiration for our work. 

\section{Model architecture}

ATGNN consists of three blocks: Patch GNN (PGN), Patch-Label GNN (PLG), and Label-Label GNN (LLG).  The PGN block models the correlation across different patches of the input mel-spectrogram. The PLG block models captures the relationship between input patches \& label embeddings and the  LLG block models the correlation between different label embeddings. Following the convention of \cite{yaoml}, we refer to the stack  of single PLG and LLG block as a Multi-Label GNN (MLG) block.  

\subsection{PGN: Patch GNN}

The input spectrogram with size $ F \times T \times 1$ is first divided into $N$ patches, where $F$ denotes the frequency axis dimension and $T$ denotes the time axis dimension. Instead of directly splitting and flattening images into tokens based on a linear  patch embedding layer as used in Audio Spectrogram Transformer (AST) ~\cite{ast}, a CNN based backbone is adopted for extraction of patches. Input $X \in \mathbb{R} ^{ F \times T \times 1} $ is transformed to a feature map $X_{t} \in \mathbb{R}^ {M}$, where $M = \left( F/p \ \times \ T/p \right) \times D $, with $p$ as the reduction ratio and $D$ as the feature dimension.  The resulting feature map is flattened as $\mathbb{R} ^{FT/p^{2} \times D}$ and summed with a learnable positional encoding. The flattened feature map can be viewed as a set of unordered nodes from which a $k$-nearest neighbour graph is constructed based on Euclidean distance between nodes. 
The graph is further updated by a graph convolution layer, which can exchange information between neighbouring nodes through message passing operation. Specifically, node $x_{i}$ is updated using max-relative graph convolution~\cite{li2019deepgcns}:
\begin{equation}
g(\cdot) = x^{''} = [x_{i}, \max(x_{j} - x_{i}) \ | \ j \in \textit{N}(x_{i})]
\end{equation}
\begin{equation}
h(\cdot) = x^{'} = x^{''}W_{update},
\end{equation}
where $N(x_{i})$ are the neighbours of the node $x_{i}$ and $x^{'}$ and  $x^{''}$ denote updated node embeddings through different update operations. Combination of (1) and (2) is termed as \textit{GraphConv}. A linear layer is applied to each node before and after GraphConv to increase feature diversity and avoid oversmoothing, followed by nonlinear activation. The updated graph convolution operation can be denoted by 
\begin{equation}
     y_{i} = \sigma(GraphConv(W_{in} x_{i}))W_{out} + x_{i},
\end{equation}
where $y_{i}$ denotes the updated node embedding, $\sigma$ denotes non linearity, $W_{in}$ and $W_{out}$ are weights of fully connected layers applied before and after \textit{GraphConv}. Similar to Transformer models~\cite{vaswani2017attention}, a feed-forward layer network (FFN) is applied to each node embedding. FFN essentially comprises two linear layers with a non-linear activation between the layers. The combination of  GraphConv followed by  FFN layer comprises a single PGN block. 

In order to stack multiple PGN layers and avoid oversmoothing due to depth, dilated aggregation is adopted for the \textit{GraphConv} operation.  Dilated convolutions were proposed in \cite{yu2015multi} as an alternative to max pool operations. Specifically, for an
input graph $G = (V, E)$ with dilated $k$-NN and $d$ as the dilation rate, the dilated $k$-NN returns the $k$ nearest neighbors within the $k \times d$ neighborhood region by skipping every $d$
neighbors. Dilated aggregation helps in maintaining feature diversity and reduces the over-smoothing across graph layers.

The PGN block  architecture has two flavours, namely isotropic and pyramid~\cite{yaoml}. Isotropic architecture is used by Transformer styled models where the feature dimension remains constant across layers whereas pyramid architecture is primarily used in CNN based architectures like ResNet~\cite{he2016deep}, where feature maps are down-sampled across subsequent layers. 

\linesubsec{Isotropic architecture} The number of nodes for this architecture is set to  $N = FT/256$, depending on the value of $F$ and $T$.  The number of $k$ in $k$-NN is linearly increased from $k$  to $2 \cdot k$ across the layers. 

\linesubsec{Pyramid architecture} 
%Pyramids are like like stacks of filtered inputs with exponentially reduced dimensions. 
The pyramid architecture is used to generate multi-scale feature maps to exploit the compositional hierarchies of the input, where higher level features are obtained by composition of local level features captured in the initial layers. 
In addition to the learnable positional embedding, a relative positional encoding similar to \cite{liu2021swin} is used in the pyramid model. For node $i$ and $j$, if the positional encoding is $e_{i} $ and $e_{j}$, then their relative positional distance between them is $e_{i}^\mathsf{T} e_{j}$. This distance is added to the feature embedding distance to construct the $k$-NN graph~\cite{VIG-GNN}.

\subsection{PLG: Patch-Label GNN}

The PLG block maps the correlation between patch and label embeddings, where patch embeddings $X = \{x_{1},x_{2}, \ldots, x_{N} \} \in \mathbb{R}^{N \times C}$ are the output of $M$ PGN blocks and $L = \{l_{1},l_{2}, \ldots, l_{S}\} \in \mathbb{R}^{S \times C} $ are the learnable label embeddings with dimensionality $C$ and $S$ number of classes. Each label node connects to its $k_{plg}$ patch node neighbours measured by Euclidean distance, post which max-relative graph convolution is used to update the label nodes as follows:
%Note that the output of PGN blocks have different feature dimension than the label embeddings, hence we use a linear projection layer to project the patch features from dimension $D$ to $S$ in order to match the label embedding dimension.  
\begin{equation}
    l_{i}^{''} = [l_{i}, \max(l_{i} - x{j}) \ | \ j \in N(l_{i}) ] 
\end{equation}
\begin{equation}
       l_{i}^{'} = l_{i} + l_{i}^{''}W_{l-update}
\end{equation}
where $l_{i}$ is the $i$-th label node and $W_{l-update} \in \mathbb{R} ^{S \times S}$ is the learnable update matrix for label nodes.  The label embeddings are updated by message passing from the corresponding label nodes, which builds the correlation between the spectrogram regions and the label embeddings. 

\subsection{LLG: Label-Label GNN}

The LLG block is used to map correlation between different label embeddings. The updated label embeddings from PLG blocks are used as nodes in a fully connected graph and each node is updated by aggregating information from the other nodes in the following manner:
\begin{equation}
    \hat{L} = AL^{'} + L^{'},
\end{equation}
where $L^{'} = [l_1 ^{'},l_2 ^{'},\ldots, l_S ^{'}] \in R ^{S \times C}$ are updated label embeddings from the PLG block. $A \in R^{S \times S}$ is a learnable adjacency matrix with random initialisation. $\hat{L} =  [\hat{l_1},\hat{l_2},\ldots, \hat{l_S}] \in R ^{S \times C}$ is the matrix of refined label node embeddings. $A$ is learnt during training of the model and captures the latent label correlations.  
%The output of the LLG block is fed into a FFN network similar to the PLG block in order to map the non-linear interaction between the label nodes. 

\subsection{Prediction}

The final output of the PGN block is aggregated via global average pooling and a set of $1 \times 1$ convolution layers with nonlinear activation are applied to obtain the final patch logits:
\begin{equation}
    Y_{patch} = Conv_{p}(AvgPool(X_{patch}) ),
\end{equation} 
%\newline
where $Y_{patch}\!\in\!\mathbb{R}^{1 \times S}$ denotes the final patch logits. $Conv_{p}$ denotes the set of $1\!\times\!1$ convolution layer, $AvgPool$ denotes the average pooling operation applied on the final output $X$of the PGN block. 

For label nodes,  a readout function $R$ projects each of the $S$ label embeddings  $l_i ^{'}$ using a projection matrix $W ^{p} \in \mathbb{R}^ {S \times d}$  where $W^{p}_i$ is the learned output vector for $l^{'}_i$. This is akin to applying a separate linear layer to map the presence of each label in the given input: 
\begin{equation}
     \hat{y}_{i} = W^{p}_i l_i^{'\mathsf{T}}.
\end{equation}
The classification score of each label embedding are concatenated as $ \hat{Y}\!=\![\hat{y}_1,\hat{y}_2,\ldots,\hat{y}_S]$ $\in \mathbb{R}^{1\times S}$. The final score is obtained as:
\begin{equation}
    Y = sigmoid(Y_{patch} + \hat{Y}).
\end{equation}

\section{Experiments}

\subsection{Datasets}

We evaluated the proposed models on two commonly used datasets: AudioSet~\cite{AUDIOSET} and FSD50K~\cite{fsd50k}. 
AudioSet is a large scale weakly labelled  audio event dataset 
of over 2 million 10-second audio clips extracted from YouTube
videos. The audio events in the datasets are categorised into 527 predefined labels with each audio clip containing one or more audio events. We use the balanced subset of this dataset comprising 20550 training samples and 1887 evaluation samples. 
FSD50K~\cite{fsd50k} is a publicly available weakly labelled  dataset comprising sound event audio clips categorized across  200 classes drawn from the AudioSet ontology.  The dataset comprises 3 subsets: training, validation and evaluation subset consisting of 37134, 4170 and 10231  samples respectively. The audio clips in the  dataset are of variable length, ranging from 0.3 to 30s.  

\subsection{Training pipeline}

In order to conduct a fair evaluation, we adopted the training pipeline from \cite{PSLA}, details of which are as follows:

\linesubsec{Data Preprocessing} The audio files were first resampled to 16 kHz and a short-time Fourier transform (STFT) with a window size of 25ms and hop length of 10 ms,  was applied to each audio clip in order to compute a spectrogram. Subsequently, a 128-dimensional mel filter bank was applied followed by a logarithmic operation to extract its log-mel spectrogram. Due to variability in audio clip duration, we applied zero-padding to achieve sample-wise uniform length of 1024 frames for FSD50K and 1056 for AudioSet experiments. \\

\linesubsec{Model details}  The PGN block of the proposed  models was initialised with ImageNet pre-trained weights from~\cite{VIG-GNN} for all the experiments. The MLG  blocks were trained from scratch. The terms ``iso" and ``pyr" respectively denote isometric and pyramid versions of the PGN block. The pyramid model has two different sizes - small (s) and medium (med). In case of pyramid model, a stage-wise approach was followed where each stage consists of $M$ PGN blocks followed by $P$ MLG blocks. In case of pyr-s, we found the best results with $M = [2,2,6,2]$ and $P = [1,1,3,1]$. For pyr-med, best results were achieved with  $M = [2,2,16,2]$ and $P = [1,1,6,1]$. In our experiments, we varied $k$ from 9 to 20 as shown in Table~\ref{tab:table4} for the PGN block and observed that the $k$ values did not change the results significantly, hence for all experiments we used  $k$ and $k_{plg}$ $=9$.   
     
\begin{table}[t]
   \caption{Mean Average Precision vs $k$-NN values for PGN block}
  \centering
    \begin{tabular}{l c c c c c} \toprule
     \textbf{k} & \textbf{9} &\textbf{12}
     &\textbf{15} &\textbf{18} &\textbf{20}\\ 
     \midrule
      mAP & 57.9 & 57.7 & 57.7 & 57.9 & 57.6\\ 
      \bottomrule
    \end{tabular}
    \label{tab:table4}

\end{table}

\linesubsec{Balanced Sampling}  We adopted a  random balanced sampling approach for FSD50K, where each audio clip is assigned a sampling weight such that higher weight is assigned to clips containing rare events. For more details, refer to~\cite{PSLA}.  

\linesubsec{Data Augmentation} We used mixup~\cite{zhang2017mixup} data augmentation (mixup ratio $= 0.5$)  and time-frequency masking ~\cite{specaugment} with maximum time mask of 192 frames and maximum frequency mask of 48 bins for all the experiments. 

\linesubsec{Label Enhancement} Label noise is prevalent in both FSD50K and AudioSet, hence we adopted the  enhanced labels   proposed in \cite{PSLA}.
%, where first a teacher model was trained on the full training set of AudioSet, after which a label modification threshold is calculated by  taking the mean of the prediction score from the teacher model for all audio clips originally labelled as the corresponding class.  Based on the threshold, additional labels were added to samples where the  teacher model’s prediction score of the class is greater than the corresponding label modification
%threshold. 

\linesubsec{Training details} We used an initial learning rate of 5e-4 and linear 
learning rate warm-up strategy for the first 1,000 iterations.  The learning rate was halved every 5
epochs after the 10th epoch for FSD50K experiments and after every 5 epochs after the 35th and 10th epoch for the AudioSet experiments. 
All  models were trained for 50 epochs on FSD50K and 60 epochs on AudioSet with batch size of 24 along  with the Adam optimizer~\cite{kingma2014adam} and binary cross-entropy. 

\section{Results and Discussion}

In this section we present the results of our experiments. Tables~\ref{tab:table1} and \ref{tab:table2} showcase the results obtained on the evaluation set of AudioSet balanced and FSD50k. We primarily compare ATGNN with AST and PSLA models, both of which have achieved SOTA results on AudioSet and FSD50k. 
It can be seen from Table~\ref{tab:table1} that the isometric version of ATGNN achieves comparable performance to AST and outperforms the PSLA model, wheres both the pyramid versions outperform AST model. Our results for AST differs from the results reported in the original paper~\cite{ast} due to mismatch in our training set size with~\cite{PSLA}, hence it was not possible to use the label enhancement files provided by~\cite{PSLA}.

\begin{table}[]
    \centering
    \caption{Results on AudioSet-balanced. $\ast$ indicates our run.}
    \begin{tabular}{l c c} \toprule
        \textbf{Model} & \textbf{\#Params} &  \textbf{mAP} \\ \midrule
        PANNS~\cite{kong2020panns}  & 81.0\,M & 0.278\\
         PSLA~\cite{PSLA} $\ast$ &  13.6\,M & 0.308 \\
         AST~\cite{ast} $\ast$ & 88.7\,M & 0.330 \\ \midrule
         ATGNN-iso$\ast$ & 38.7\,M & 0.330 \\
         ATGNN-pyr-s$\ast$ & 36.4\,M & 0.335  \\
         ATGNN-pyr-s (-MLG)$\ast$ & 36.4\,M & 0.331  \\
         ATGNN-pyr-med$\ast$ & 62.7\,M & \textbf{0.336}  \\ 
          ATGNN-pyr-med (-MLG)$\ast$ & 62.7\,M & 0.332 \\ \bottomrule
    \end{tabular}
    \label{tab:table1}
\end{table}
 
%\begin{table}[H]
%    \captionsetup{singlelinecheck = false, format= hang, justification=raggedright}
%   \caption{Results on AudioSet-balanced. $\ast$ indicates our run. }
%    \begin{tabular}{ccc}
%     \hline
%         \textbf{Model} & \textbf{\#Params} &  \textbf{mAP}  \\
%         \hline
%        
%         PANNS~\cite{kong2020panns}  & 81M & 0.278\\
%        
%         PSLA ~\cite{PSLA} $\ast$ &  13.64M & 0.308 \\
%        
%         AST ~\cite{ast} $\ast$ & 88.7M & 0.3301 \\
%         
%         SpecGNN-iso$\ast$ & 38.7M & 0.330 \\
%         
%         SpecGNN-pyr-s$\ast$ & 36.4M & 0.3351 \\
%         
%         SpecGNN-pyr-med$\ast$ & 62.7M & 0.3367 \\
%    \end{tabular}
%    
%    \label{tab:table1}
%\end{table}

\begin{table}[]
    \centering
    %\captionsetup{singlelinecheck = false, format= hang, justification=raggedright}
    \caption{Results on FSD50K. $\ast$ indicates our run.}
    \begin{tabular}{l c c} \toprule
         \textbf{Model} & \textbf{\#Params} & \textbf{ mAP}  \\\midrule
        
         FSD50K Baseline~\cite{fsd50k}  & 0.27M  & 0.434\\
        
        Wav2CLIP~\cite{wu2022wav2clip} & - & 0.431 \\
        
        Audio Transformers~\cite{verma2021audio} & 2.3M & 0.537\\
        
        PSLA~\cite{PSLA}$\ast$  &  13.6M & 0.559 \\
        
         AST~\cite{ast}$\ast$ & 88.7M & 0.572 \\ \midrule
         
         ATGNN-iso$\ast$  & 38.7M & 0.570 \\
         
         ATGNN-pyr-s$\ast$  & 36.4M & 0.583 \\
         ATGNN-pyr-s (-MLG)$\ast$  & 36.4M & 0.579 \\
         ATGNN-pyr-med$\ast$  & 62.7M & \textbf{0.585} \\ 
         ATGNN-pyr-med$\ast$(-MLG)  & 62.7M & 0.580 \\ \bottomrule
    \end{tabular}
    \label{tab:table2}
\end{table}

The results are similar for the FSD50K dataset as well where the isometric architecture is competent with AST and the pyramid architecture outperforms PSLA and AST. As shown in Tables~\ref{tab:table1} and \ref{tab:table2}, the MLG block brings an additional benefit of $\approx 0.4$ mAP to the overall score, implying that the SGN block can be used as a standalone model for obtaining comparable results to the SOTA models for audio classification tasks.

\section{Conclusions}
%We presented an end-to-end GNN model that learns feature and label correlations for audio tagging. This model extracts local features through a CNN backbone and maps global context through graph convolution operations. The model also learns cross-correlation across label and feature embeddings in an end-end fashion and consistently outperforms SOTA models across two widely used audio classification benchmarks. We intend to expand this work beyond a patch based graph construction approach and adopt  classical methods like visibility graphs for time series or consider more unexplored pathways such as reformulating the spectrogram as a 3D-point cloud. Another interesting direction would be to consider multi-range edges for diverse spatial interactions such as local edges for fine-grained relations and  long range edges to incorporate signal harmonics. Finally, through this work, we hope that graph based models  can be considered as an alternative representation learning method in addition to convolution and attention based architectures for audio classification tasks. 

We presented an end-to-end GNN model that learns feature and label correlations for audio tagging. This model uses a CNN for local feature extraction, and graph convolution operations for global context mapping. Crucially, it learns cross-correlations between label and feature embeddings, consistently surpassing state-of-the-art models on two major audio classification benchmarks. Future work will explore beyond patch-based graph construction, considering classical methods like visibility graphs for time series or innovative concepts like treating the spectrogram as a 3D-point cloud. Additionally, the adoption of multi-range edges could enable nuanced spatial interactions, with local edges for detailed relations and long-range edges for signal harmonics.We hope our work encourages consideration of graph-based models as alternate representation learning methodologies, alongside convolution and attention-based architectures, for audio classification tasks.

%We've proposed an end-to-end Graph Neural Network (GNN) model for audio classification that uniquely learns correlations between features and labels. This model uses a Convolutional Neural Network (CNN) for local feature extraction, and graph convolution operations for global context mapping. Crucially, it learns cross-correlations between label and feature embeddings, consistently surpassing state-of-the-art models on two major audio classification benchmarks.

%Future work will explore beyond patch-based graph construction, considering classical methods like visibility graphs for time series or innovative concepts like treating the spectrogram as a 3D-point cloud. Additionally, the adoption of multi-range edges could enable nuanced spatial interactions, with local edges for detailed relations and long-range edges for signal harmonics.

%We hope our work encourages consideration of graph-based models as alternate representation learning methodologies, alongside convolution and attention-based architectures, for audio classification tasks.

\bibliographystyle{IEEEtran}
\bibliography{mybib}

\begin{thebibliography}{10}
\providecommand{\url}[1]{#1}
\csname url@samestyle\endcsname
\providecommand{\newblock}{\relax}
\providecommand{\bibinfo}[2]{#2}
\providecommand{\BIBentrySTDinterwordspacing}{\spaceskip=0pt\relax}
\providecommand{\BIBentryALTinterwordstretchfactor}{4}
\providecommand{\BIBentryALTinterwordspacing}{\spaceskip=\fontdimen2\font plus
\BIBentryALTinterwordstretchfactor\fontdimen3\font minus
  \fontdimen4\font\relax}
\providecommand{\BIBforeignlanguage}[2]{{%
\expandafter\ifx\csname l@#1\endcsname\relax
\typeout{** WARNING: IEEEtran.bst: No hyphenation pattern has been}%
\typeout{** loaded for the language `#1'. Using the pattern for}%
\typeout{** the default language instead.}%
\else
\language=\csname l@#1\endcsname
\fi
#2}}
\providecommand{\BIBdecl}{\relax}
\BIBdecl

\bibitem{hershey2017cnn}
S.~Hershey, S.~Chaudhuri, D.~P. Ellis, J.~F. Gemmeke, A.~Jansen, R.~C. Moore,
  M.~Plakal, D.~Platt, R.~A. Saurous, B.~Seybold \emph{et~al.}, ``Cnn
  architectures for large-scale audio classification,'' in \emph{2017 ieee
  international conference on acoustics, speech and signal processing
  (icassp)}.\hskip 1em plus 0.5em minus 0.4em\relax IEEE, 2017, pp. 131--135.

\bibitem{ast}
Y.~Gong, Y.-A. Chung, and J.~Glass, ``Ast: Audio spectrogram transformer,''
  \emph{arXiv preprint arXiv:2104.01778}, 2021.

\bibitem{koutini2021efficient}
K.~Koutini, J.~Schl{\"u}ter, H.~Eghbal-zadeh, and G.~Widmer, ``Efficient
  training of audio transformers with patchout,'' \emph{arXiv preprint
  arXiv:2110.05069}, 2021.

\bibitem{liu2023efficientvit}
X.~Liu, H.~Peng, N.~Zheng, Y.~Yang, H.~Hu, and Y.~Yuan, ``Efficientvit: Memory
  efficient vision transformer with cascaded group attention,'' in
  \emph{Proceedings of the IEEE/CVF Conference on Computer Vision and Pattern
  Recognition}, 2023, pp. 14\,420--14\,430.

\bibitem{chen2019multi}
Z.-M. Chen, X.-S. Wei, P.~Wang, and Y.~Guo, ``Multi-label image recognition
  with graph convolutional networks,'' in \emph{Proceedings of the IEEE/CVF
  conference on computer vision and pattern recognition}, 2019, pp. 5177--5186.

\bibitem{shrivastava2020mt}
H.~Shrivastava, Y.~Yin, R.~R. Shah, and R.~Zimmermann, ``Mt-gcn for multi-label
  audio-tagging with noisy labels,'' in \emph{ICASSP 2020-2020 IEEE
  International Conference on Acoustics, Speech and Signal Processing
  (ICASSP)}.\hskip 1em plus 0.5em minus 0.4em\relax IEEE, 2020, pp. 136--140.

\bibitem{VIG-GNN}
K.~Han, Y.~Wang, J.~Guo, Y.~Tang, and E.~Wu, ``Vision gnn: An image is worth
  graph of nodes,'' in \emph{NeurIPS}, 2022.

\bibitem{yaoml}
R.~Yao, S.~Jin, W.~Liu, C.~Qian, P.~Luo, and J.~Wu, ``Ml-vig: Multi-label image
  recognition with vision graph convolutional network.''

\bibitem{liu2021time}
J.~Liu, Y.~Song, L.~Wang, J.~Dang, and R.~Yu, ``Time-frequency representation
  learning with graph convolutional network for dialogue-level speech emotion
  recognition.'' in \emph{Interspeech}, 2021, pp. 4523--4527.

\bibitem{kwon2022multi}
Y.~Kwon, H.-S. Heo, J.-w. Jung, Y.~J. Kim, B.-J. Lee, and J.~S. Chung,
  ``Multi-scale speaker embedding-based graph attention networks for speaker
  diarisation,'' in \emph{ICASSP 2022-2022 IEEE International Conference on
  Acoustics, Speech and Signal Processing (ICASSP)}.\hskip 1em plus 0.5em minus
  0.4em\relax IEEE, 2022, pp. 8367--8371.

\bibitem{wang2020speaker}
J.~Wang, X.~Xiao, J.~Wu, R.~Ramamurthy, F.~Rudzicz, and M.~Brudno, ``Speaker
  diarization with session-level speaker embedding refinement using graph
  neural networks,'' in \emph{ICASSP 2020-2020 IEEE International Conference on
  Acoustics, Speech and Signal Processing (ICASSP)}.\hskip 1em plus 0.5em minus
  0.4em\relax IEEE, 2020, pp. 7109--7113.

\bibitem{sun2020ontology}
Y.~Sun and S.~Ghaffarzadegan, ``An ontology-aware framework for audio event
  classification,'' in \emph{ICASSP 2020-2020 IEEE International Conference on
  Acoustics, Speech and Signal Processing (ICASSP)}.\hskip 1em plus 0.5em minus
  0.4em\relax IEEE, 2020, pp. 321--325.

\bibitem{li2019deepgcns}
G.~Li, M.~Muller, A.~Thabet, and B.~Ghanem, ``Deepgcns: Can gcns go as deep as
  cnns?'' in \emph{Proceedings of the IEEE/CVF international conference on
  computer vision}, 2019, pp. 9267--9276.

\bibitem{vaswani2017attention}
A.~Vaswani, N.~Shazeer, N.~Parmar, J.~Uszkoreit, L.~Jones, A.~N. Gomez,
  {\L}.~Kaiser, and I.~Polosukhin, ``Attention is all you need,''
  \emph{Advances in neural information processing systems}, vol.~30, 2017.

\bibitem{yu2015multi}
F.~Yu and V.~Koltun, ``Multi-scale context aggregation by dilated
  convolutions,'' \emph{arXiv preprint arXiv:1511.07122}, 2015.

\bibitem{he2016deep}
K.~He, X.~Zhang, S.~Ren, and J.~Sun, ``Deep residual learning for image
  recognition,'' in \emph{Proceedings of the IEEE conference on computer vision
  and pattern recognition}, 2016, pp. 770--778.

\bibitem{liu2021swin}
Z.~Liu, Y.~Lin, Y.~Cao, H.~Hu, Y.~Wei, Z.~Zhang, S.~Lin, and B.~Guo, ``Swin
  transformer: Hierarchical vision transformer using shifted windows,'' in
  \emph{Proceedings of the IEEE/CVF international conference on computer
  vision}, 2021, pp. 10\,012--10\,022.

\bibitem{AUDIOSET}
J.~F. Gemmeke, D.~P. Ellis, D.~Freedman, A.~Jansen, W.~Lawrence, R.~C. Moore,
  M.~Plakal, and M.~Ritter, ``Audio set: An ontology and human-labeled dataset
  for audio events,'' in \emph{2017 IEEE international conference on acoustics,
  speech and signal processing (ICASSP)}.\hskip 1em plus 0.5em minus
  0.4em\relax IEEE, 2017, pp. 776--780.

\bibitem{fsd50k}
E.~Fonseca, X.~Favory, J.~Pons, F.~Font, and X.~Serra, ``Fsd50k: an open
  dataset of human-labeled sound events,'' \emph{IEEE/ACM Transactions on
  Audio, Speech, and Language Processing}, vol.~30, pp. 829--852, 2021.

\bibitem{PSLA}
\BIBentryALTinterwordspacing
Y.~Gong, Y.-A. Chung, and J.~Glass, ``{PSLA}: Improving audio tagging with
  pretraining, sampling, labeling, and aggregation,'' \emph{{IEEE}/{ACM}
  Transactions on Audio, Speech, and Language Processing}, vol.~29, pp.
  3292--3306, 2021. [Online]. Available:
  \url{https://doi.org/10.1109%2Ftaslp.2021.3120633}
\BIBentrySTDinterwordspacing

\bibitem{zhang2017mixup}
H.~Zhang, M.~Cisse, Y.~N. Dauphin, and D.~Lopez-Paz, ``mixup: Beyond empirical
  risk minimization,'' \emph{arXiv preprint arXiv:1710.09412}, 2017.

\bibitem{specaugment}
D.~S. Park, W.~Chan, Y.~Zhang, C.-C. Chiu, B.~Zoph, E.~D. Cubuk, and Q.~V. Le,
  ``Specaugment: A simple data augmentation method for automatic speech
  recognition,'' \emph{arXiv preprint arXiv:1904.08779}, 2019.

\bibitem{kingma2014adam}
D.~P. Kingma and J.~Ba, ``Adam: A method for stochastic optimization,''
  \emph{arXiv preprint arXiv:1412.6980}, 2014.

\bibitem{kong2020panns}
Q.~Kong, Y.~Cao, T.~Iqbal, Y.~Wang, W.~Wang, and M.~D. Plumbley, ``Panns:
  Large-scale pretrained audio neural networks for audio pattern recognition,''
  \emph{IEEE/ACM Transactions on Audio, Speech, and Language Processing},
  vol.~28, pp. 2880--2894, 2020.

\bibitem{wu2022wav2clip}
H.-H. Wu, P.~Seetharaman, K.~Kumar, and J.~P. Bello, ``Wav2clip: Learning
  robust audio representations from clip,'' in \emph{ICASSP 2022-2022 IEEE
  International Conference on Acoustics, Speech and Signal Processing
  (ICASSP)}.\hskip 1em plus 0.5em minus 0.4em\relax IEEE, 2022, pp. 4563--4567.

\bibitem{verma2021audio}
P.~Verma and J.~Berger, ``Audio transformers: Transformer architectures for
  large scale audio understanding. adieu convolutions,'' \emph{arXiv preprint
  arXiv:2105.00335}, 2021.

\end{thebibliography}
\end{document}